\documentstyle[preprint,aps]{revtex}
\begin{document}
\draft

\title{Pion photoproduction in the delta resonance region}
\author{R.A. Arndt, I.I. Strakovsky and R.L. Workman}
\address{Department of Physics, Virginia Polytechnic Institute and State
University, Blacksburg, VA 24061}

\date{\today}
\maketitle

\begin{abstract}
We have updated our multipole analyses to incorporate new data
from the low-energy and delta resonance regions. We note a slight
decrease in our estimate of the delta photo-decay amplitudes. This
agrees with results found in Compton scattering. 
Our values are also in good agreement with a previous RPI determination.
We have reexamined our determinations of the E2/M1 ratio at the resonance
energy and at the pole. We find an E2/M1 ratio (at the pole) which is in 
good agreement with recent the Mainz value. 

\end{abstract}

 
\narrowtext
\section{INTRODUCTION}

Recent interest in the pion photoproduction reaction has been focused on
the threshold and delta resonance regions. Most studies have been motivated 
by a desire to check the predictions of chiral perturbation theory and to 
pin down the photo-decay amplitudes of the
$\Delta (1232)$. With the flood of new and precise data in this region, 
we have reexamined the $\Delta(1232)$ photo-decay amplitudes coming from our 
analyses of the pion photoproduction data base. Given the recent interest in
a more precise determination of the E2/M1 ratio, we have extracted this
quantity as well. 

Our method of analysis is the same as was used to generate the
results of Ref.\cite{vpi96}, with one modification.  
The $\pi$N interaction is currently being 
determined by the pion C.M. momentum rather than the C.M. energy. This 
gives a proper threshold behavior for all charge-channel
multipoles but shifts the on-shell resonance points (in W$_{C.M.}$)
slightly according to the specific charge channel being considered. The
pole position is not altered.

\section{THE DELTA PHOTO-DECAY AMPLITUDES}     

Most important to the present analysis were the recent total cross section
measurements of Ref.\cite{mac}. Our previous analysis\cite{vpi96} predicted
total $\pi^0 p$ and $\pi^+ n$ total cross sections systematically above
these new data. This was also suggested in an analysis of Compton scattering
data\cite{compton}. In that work, a $2.8\pm 0.9$3\% reduction in our 
$M^{3/2}_{1+}$ amplitude was found to be necessary. 

Our present findings complement, and now agree with, the value found in  
Ref.\cite{compton}. For the helicity 1/2 and 3/2 photo-decay 
amplitudes, we find values about 4\% lower than our previous determinations.
These are compared to values from the Glasgow\cite{Crawford} and RPI\cite{rpi}
groups in Table~I. Though the RPI values have larger errors, the central 
values from the VPI and RPI\cite{rpi} determinations are in good agreement. 
The ranges for A$_{1/2}$ and  A$_{3/2}$  were chosen to cover 
results found in fits to our global solution to 2~GeV (SP97),  
a more restricted analysis to 500 MeV (W500), and fits to our 
single-energy solutions (SES)\cite{said}.

\section{THE E2/M1 RATIO}

There are a number of ways to define the E2/M1 ratio. In the following, 
we will restrict 
ourselves to determinations which can be made directly from the multipole
amplitudes. The resonance value which we quote is determined by evaluating
the ratio Im$E^{3/2}_{1+}$/Im$M^{3/2}_{1+}$ at the resonance energy. 
Both the RPI\cite{rpi} and VPI\cite{vpi92} groups have found ratios near 
$-1.5\%$ in the past. However, while the PDG\cite{pdg} quotes these values, 
a number of recent determinations have found larger negative values. 

Khandaker and Sandorfi\cite{Sandorfi} have suggested that a higher value is
required to fit the new LEGS beam-asymmetry ($\Sigma$) data\cite{pc}. 
This claim is
supported in a recent paper from the Mainz group\cite{Beck}. A value 
consistent with the finding of Ref.\cite{Sandorfi} 
was found in an analysis of the new Mainz\cite{Beck} cross section and 
$\Sigma$ data. A second Mainz analysis\cite{Hanstein} used dispersion
relations in analyzing data over the delta resonance region. In this work,
the speed-plot method was used to determine an E2/M1 ratio at the pole.

In Table~II, we compare our updated values for the E2/M1 ratio at the 
resonance and the pole with both recent and older determinations. We find 
good agreement with the Mainz pole value, supporting the view that
this quantity is relatively model-independent. Adding the new Mainz data 
did not increase the E2/M1 ratio as much as was found in Ref.\cite{Beck}. 
We attribute this to differing assumptions and data bases used in the 
analyses. 

The SES results favor a slightly larger E2/M1 ratio  
than was found in our energy-dependent fits. Depending on how the data are 
binned (we compared results using 10 MeV and 25 MeV bins), the SES value 
ranges between approximately $-1.5$\% and $-1.9$\%. 
This value is closer to the one found in Ref.\cite{Beck}. However, 
the low-energy analysis (W500), 
which results in a very different E2/M1 ratio ($-1.2\%$),  fits the 
Mainz cross section and $\Sigma$ data\cite{Beck} quite well. 
(The overall $\chi^2$/data is 193/182.) This variability in the E2/M1 ratio 
is greatly reduced in the ratio of pole residues. 

While we found good agreement
for the pole positions and moduli of the residues using the speed plot method,
we also determined the pole value for E2/M1 using a different method. Writing
our amplitudes in the form
\begin{equation}
M \; = \; A_B(\; 1 \; + i T_{\pi N} \;) \; + \; A_R T_{\pi N} ,
\end{equation}
where $T_{\pi N}$ is the associated $\pi N$ T-matrix, we fit $T_{\pi N}$ 
using analytic forms and extrapolated to the pole. ($A_B$ and $A_R$ are 
analytic functions determined in the energy-dependent analysis.) 
The agreement between these different methods gives further confidence 
in the pole value. A 
comparison with the Mainz value\cite{Hanstein} is given in Table~II. 
The modulus of this ratio varied by only about 10\% in our fits to SP97,
W500, and the SES. 

\section{SUMMARY AND CONCLUSIONS}
\label{sec:sum}

Our updated $\Delta (1232)$ photo-decay amplitudes now agree with both the  
results from Compton scattering and an earlier RPI determination. We also
agree with the pole value for the E2/M1 ratio found by the Mainz group. Our
determinations confirm the view that this ratio is relatively 
model-independent. Given our agreement with the Mainz pole value, our 
slightly lower value at the resonance point remains to be explained. 

There are two tests which might help to clarify the discrepancy. 
First, the pole value found in 
Ref.\cite{Hanstein} could be extrapolated to the resonance 
point\cite{INT}. Second,
the multipoles which will come from a Mainz fit to $\gamma p\to p\pi^0$ and
$\gamma p\to n\pi^+$ data (used to determine E2/M1 at the 
resonance point\cite{Beck}) 
could be `speed plotted' to determine a pole value for this ratio.   
This would give three independent sets of both pole and resonance values. 
If the extrapolation to the pole is indeed model-independent (as it appears
to be), it would be interesting to see whether amplitudes giving an E2/M1 
ratio near $-2.5$\% (at the resonance point) extrapolate to a pole value 
close to that found here and in Ref.\cite{Hanstein}.

\acknowledgments
 
We thank A. L'vov, R. Beck, and L. Tiator for helpful communications. 
This work was supported in part by a U.S. Department of
Energy Grant DE-FG05-88ER40454.

\eject

Table~I. Comparison of recent determinations of the $\Delta (1232)$
decay amplitudes. Values have the units $10^{-3}$ GeV$^{-1/2}$. See
the text for a discussion of the range of values quoted for the
VPI analyses. 
\vskip 10pt
\centerline{
\vbox{\offinterlineskip
\hrule
\halign{\hfill#\hfill&\qquad\hfill#\hfill&\qquad\hfill#\hfill\cr
\noalign{\vskip 6pt}
Source          &A$_{1/2}$  &A$_{3/2}$   \cr
\noalign{\vskip 6pt}
\noalign{\hrule}
\noalign{\vskip 10pt}
VPI (old)\cite{vpi96}     & -141$\pm$5     &  -261$\pm$5   \cr
\noalign{\vskip 6pt}
Glasgow\cite{Crawford}   & -145$\pm$15     &  -263$\pm$26   \cr
\noalign{\vskip 6pt}
RPI\cite{rpi}            & -135$\pm$16     &  -251$\pm$33   \cr 
\noalign{\vskip 6pt}
VPI                       & -135$\pm$5     &   -250$\pm$8  \cr
\noalign{\vskip 10pt}}
\hrule}}
\vfill
\eject
Table~II. Comparison of recent values for the E2/M1 ratio evaluated at the
pole (P) and resonance (R) positions.  See the text for a discussion of the
range of values quoted for the VPI analyses. 
\vskip 10pt
\centerline{
\vbox{\offinterlineskip
\hrule
\halign{\hfill#\hfill&\qquad\hfill#\hfill&\qquad\hfill#\hfill\cr
\noalign{\vskip 6pt}
Source            & Location    & Value       \cr
\noalign{\vskip 6pt}
\noalign{\hrule}
\noalign{\vskip 10pt}
RPI\cite{rpi}        &  R     & -1.57$\pm$0.72\%   \cr
\noalign{\vskip 6pt}
VPI (old)\cite{vpi92} &  R     &  -1.5$\pm$0.5\%  \cr
\noalign{\vskip 6pt}
BNL\cite{Sandorfi}    &  R  &  -2.7\% \cr
\noalign{\vskip 6pt}
Mainz\cite{Beck}      &  R  &  -2.5$\pm 0.2\pm 0.2$\% \cr
\noalign{\vskip 6pt}
VPI                   &  R     &  -1.5$\pm$0.5\%  \cr
\noalign{\vskip 6pt}
\noalign{\hrule}
\noalign{\vskip 6pt}
Mainz\cite{Hanstein}  &  P     &  -0.035 - i 0.046  \cr
\noalign{\vskip 6pt} 
VPI                   &  P     &  -0.034$\pm$5 - i 0.055$\pm$5  \cr
\noalign{\vskip 10pt}}
\hrule}}
\vfill


\begin{references}

\bibitem{vpi96} 
R.A. Arndt, I.I. Strakovsky, and R.L. Workman,
Phys. Rev. {\bf C} 53, 430 (1996).

\bibitem{mac} 
M. MacCormick et al., 
Phys. Rev. {\bf C} 53, 41 (1996).

\bibitem{compton} 
J. Peise et al., 
Phys. Lett. B 384, 37 (1996).

\bibitem{Crawford}
R.L. Crawford and W.T. Morton,
Nucl. Phys. {\bf B211}, 1 (1983).

\bibitem{rpi} 
R.M. Davidson, N.C. Mukhopadhyay, and R.S. Wittman,
Phys. Rev. {\bf D} 43, 71 (1991).

\bibitem{said}
These solutions can be viewed using SAID with a TELNET call to 
clsaid.phys.vt.edu with userid: said, or through our WWW site at
http://clsaid.phys.vt.edu.

\bibitem{vpi92}
R.L. Workman, R.A. Arndt, and Z. Li,
Phys. Rev. {\bf C} 46, 1546 (1992).

\bibitem{pdg}
R.M. Barnett et al.,
Phys. Rev. {\bf D} 54, 1 (1996).

\bibitem{Sandorfi}
M. Khandaker and A.M. Sandorfi,
Phys. Rev. {\bf D} 51, 3966 (1995).

\bibitem{pc}
A.M. Sandorfi, private communications.

\bibitem{Beck} 
R. Beck et al., 
Phys. Rev. Lett. {\bf 78}, 606 (1997). There appears to be a typographical
error in Eq.(4) of this paper. The signs of the M$_{1+}$ and M$_{1-}$ 
multipoles should be reversed. 

\bibitem{Hanstein}
O. Hanstein, D. Drechsel, and L. Tiator,
Phys. Lett. {\bf B385}, 45 (1996).

\bibitem{INT}
L. Tiator has claimed a value of $-2.4\%$ for the E2/M1 ratio
at the resonance energy, using the amplitudes of Ref.\cite{Hanstein}.
A comparison of the resonant and pole values from several sources was
given during the N$^*$ meeting at INT, Seattle, 1996 (World Scientific,
to be published), edited by T.-S.H.~Lee and W.~Roberts.  


\end{references}
\end{document}